\documentclass[showpacs,aps,graphicx,twocolumn]{revtex4}

\usepackage{graphicx}

\begin{document}


\title{Universal single-qubit non-adiabatic holonomic quantum gates in optomechanical system}

\author{Shan-Shan Chen$^{1,*}$,
Hao Zhang$^{1,}$\footnote{The first two authors contributed equally
to this work.}, Xue-Ke Song$^{1}$, Fu-Guo Deng$^{1,2}$, Ahmed
Alsaedi$^{2}$,  Tasawar Hayat$^{2,3}$, Haibo Wang$^{1}$, and
Guo-Jian Yang$^{1}$\footnote{Corresponding author: yanggj@bnu.edu.cn
}}

\address{$^{1}$Department of Physics, Applied Optics Beijing Area Major Laboratory,
Beijing Normal University, Beijing 100875, China\\
$^{2}$NAAM Research Group, Department of Mathematics, King Abdulaziz
University, Jeddah 21589, Saudi Arabia\\
$^{3}$Department of Mathematics, Quaid-I-Azam University, Islamabad
44000, Pakistan}

\date{\today}

\begin{abstract}
The non-adiabatic holonomic quantum computation with the advantages
of fast and robustness attracts widespread attention in recent
years. Here, we propose the first scheme for realizing universal
single-qubit gates based on an optomechanical system working with
the non-adiabatic geometric phases. Our quantum gates are robust to
the control errors and the parameter fluctuations, and have unique
functions to achieve  the quantum state transfer and entanglement
generation between cavities. We discuss the corresponding
experimental parameters and give some simulations. Our scheme may
have the practical applications in quantum computation and quantum
information processing.
\end{abstract}

 \maketitle

\section{Introduction}\label{sec1}

Quantum geometric phases
\cite{Pancharatnam1956,SimonPRL1983,Berry1984} are very important
resource for quantum computation. They have unique advantage of
robustness in quantum computation due to their global geometric
property in evolution process and thus attrack much attention from
both theoretical and experimental aspects
\cite{ZanardiPRA1999,VedralPRL2000,TongPRL2004,YiPRL2004,YiPRA2007,YiPRA2009,TondPRA2009,YiPRA2010,SongNJP2016}.
One of the important contributions in this field comes from Zanardi
and Rasetti \cite{ZanardiPRA1999}, who proposed the adiabatic
holonomic quantum computation (AHQC) by using the geometric phases.
It is showed that AHQC can be used to implement the high-fidelity
quantum gates because of its robustness to small random
perturbations of the path in parameter space and experimental
imperfection \cite{PachosMPB2001}. Following the idea of above AHQC,
several AHQC schemes based on different physical systems like
trapped ions \cite{DuanScience2001}, superconducting qubits
\cite{Faoro2003}, and semiconductor quantum dots \cite{Solinas2003},
etc., were developed. However, an adiabatic process may bring in
more decoherence due to a long evolution time, while the decoherence
will result in the decrease of fidelity. To solved this problem, the
non-adiabatic holonomic quantum computation (NHQC), such as the
early non-adiabatic geometric phase shift gate with NMR
\cite{WangPRL2001}, universal non-adiabatic geometric quantum gates
\cite{ZhuPRL2002} and, subsequently, more theoretical shemes
\cite{SjoqvistNJP2012,XU2012PRL,LongSR2014,LongPRA2014,zhouOe2015,XuePRA2015,XuPRA2015,TongPRA2016,XuPRA2017,XuePRApp2017}
and the experimental realizations
\cite{AbdumalikovNature2013,FengPRL2013,DuanNature2014,ArroyoCamejoNatcommu2014,AwschalomNatPhoton2016,PRLAwschalom2017,NatphtonicsKosaka2017,SciChinaLong2017}
of the NHQC were proposed. The investigations have confirmed  the
features of the built-in noise-resilience and less decoherence of
the NHQC.

An optomechanical system, where light and mechanical motion are
coupled by radiation pressure, is an important platform to realize,
in the systems ranging from quantum to classical ones, the quantum
effects in the content of quantum optics \cite{MAspelmeyerRMP2014}
and quantum information processing
\cite{LudwigPRL2012,StannigelPRL2012}. The fundamental study in this
field includes cooling of the mechanical resonator to its ground
state \cite{Schliesser2009,Liu201301,Peterson2016}, strong coupling
between the cavities and the mechanical resonator
 \cite{Groblacher2009,Thompson2008} and optomechanically induced transparency \cite{Weis2010,Safavi2011,LongOe2015}, etc. The
relevant application study concerns with quantum state operation
\cite{FiorePRL2011,WangPRL2012,LTianPRL2012,Palomaki2013} and the
quantum gate operation
\cite{LudwigPRL2012,StannigelPRL2012,ZhangJPB2015,AsjadOE2015}.

In this paper, we propose the first scheme to achieve a set of
universal single-qubit non-adiabatic holonomic quantum gates
(SQNAHQGs) based on an optomechanical system working with the
non-adiabatic geometric phases. This optomechanical system is
composed of two optical cavities coupling to an mechanical
oscillator, and the universal SQNAHQGs include noncommute Not gate,
phase gate and Hadamard gate, obtained in the computational basis of
the single excited state of the optomechanical system after a cyclic
evolution of the system is finished. With these universal
single-qubit gates, we can also achieve the quantum state transfer
and the entanglement generation between two cavity-modes. Our scheme
is of all the good properties of the NHQC based on a quantum system,
such as the built-in noise-resilience, faster operation, less
decoherence and non-requirement for the resource and time to remove
the dynamical phases. It provides a prototype of quantum gates
realized in the space of the mechanical motion degree of freedom,
which has the promising application in quantum computation and
quantum information processing.

This paper is organized as follows: In Sec.~\ref{sec2}, we give the
review description of the optomechanical system. In Sec.~\ref{sec3},
we show how to realize the universal single-qubit gates in the
optomechanics by using the non-adiabatic geometric phases. In
Sec.~\ref{sec4}, we give some numerical simulations and discussions.
A summary is given in Sec.~\ref{sec5}.

\begin{figure}[!ht]
\begin{center}
\includegraphics[width=6.5cm,angle=0]{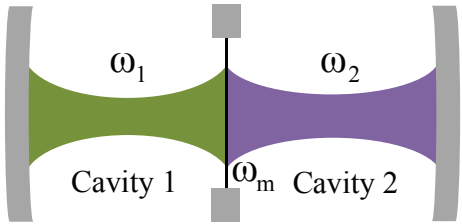}
\caption{ Schematic diagram of the optomechanical system composed of
the two cavities and a mechanical oscillator. The mechanical oscillator is realized by a membrane fixed in middle. $\omega_{1}$,
$\omega_{2}$, and $\omega_{m}$ represent the frequencies of  cavity
1, cavity 2, and the mechanical oscillator, respectively.
}\label{fig1}
\end{center}
\end{figure}

\section{Basic model for an optomechanical system}  \label{sec2}

The optomechanical system under consideration is shown by  Fig.~\ref{fig1}, where the
two cavity modes coupled to each other by radiation pressure force via a
mechanical oscillator, and also are driven respectively by a laser in the red
sideband resonant with mechanical mode. After the linearization procedure,
the Hamiltonian of this optomechanical system in the interaction picture is
given by($\hbar=1$) \cite{WangPRL2012,LTianPRL2012}
\begin{eqnarray}      \label{eqH1}
\hat{H}_{1}=\sum_{i=1,2}\delta_{i}\hat{a}_{i}^{\dag}\hat{a}_{i}
+  G_{i}\hat{a}_{i}\hat{b}_{m}^{\dag} + H.c.,
\end{eqnarray}
where $\hat{a}_{i}$ ($\hat{a}_{i}^{\dagger }$) ($i=1,2$) and $\hat{b}_{m}$ ($%
\hat{b}_{m}^{\dagger }$) are the annihilation (creation) operators for the $i
$th cavity of frequency $\omega _{i}$ and the mechanical oscillator of
frequency $\omega _{m}$, respectively. $\delta _{i}=-\Delta _{i}-\omega _{m}$
with the detuning $\Delta _{i}=\omega _{di}-\omega _{i}$ between the laser $%
\omega _{di}$ and the cavity mode $\omega _{i}$. $G_{i}=g_{0i}\sqrt{n_{i}}$ is the effective coupling
strength which depends on the single-photon coupling strength $g_{0i}$ and the
intracavity photon number $n_{i}$.

In this paper, we choose $\delta_{i}=0$. We assume that
$|g_{1}\rangle=|100\rangle$, $|g_{2}\rangle=|001\rangle$, and
$|e\rangle=|010\rangle$ represent the single excited states on
cavities 1, 2, and the mechanical oscillator, respectively. We makes
$|g_{1}\rangle$ and $|g_{2}\rangle$ as the qubit basis states and
$|e\rangle$ as the ancillary qubit to construct a single-qubit state
subspace $S_{1}=\{|g_{1}\rangle, |e\rangle, |g_{2}\rangle\}$. In
this single-excitation subspace, the Hamiltonian (\ref{eqH1}) can be rewritten as
\begin{eqnarray}      \label{eqH2}
\hat{H}_{2}=G_{0}(t)\left[\sin\frac{\theta}{2}e^{i\varphi}|e\rangle\langle
g_{1}| - \cos\frac{\theta}{2}|e\rangle\langle g_{2}| + H.c.\right],
\end{eqnarray}
where $ G_{0}(t)=\sqrt{G_{1}^{2}(t) + G_{2}^{2}(t)}$. The Rabi
frequencies $G_{1}(t)$ and $G_{2}(t)$ satisfy the ratio
$G_{1}(t)/G_{0}(t)=\sin\frac{\theta}{2}e^{i\varphi}$ and
$G_{2}(t)/G_{0}(t)=-\cos\frac{\theta}{2}$, respectively. Therefore,
the Hamiltonian (\ref{eqH2}) can be expressed in the matrix form
\begin{eqnarray}        \label{vector}
\hat{H}_{3}=G_{0}(t)\left[
\begin{array}{ccc}
0&\sin\frac{\theta}{2}e^{-i\varphi}&0\\
\sin\frac{\theta}{2}e^{i\varphi}&0&-\cos\frac{\theta}{2}\\
0&-\cos\frac{\theta}{2}&0\\
\end{array}
\right],
\end{eqnarray}
where $|g_{1}\rangle$, $|e\rangle$, and $|g_{2}\rangle$ are shown as
$|g_{1}\rangle=[1,0,0]^{T}$, $|e\rangle=[0,1,0]^{T}$, and
$|g_{2}\rangle=[0,0,1]^{T}$, respectively. The instantaneous
eigenvectors of the Hamiltonian (\ref{vector})  are given by
\begin{eqnarray}      \label{eqH20}
|E_{0}\rangle& = &\cos\frac{\theta}{2}|g_{1}\rangle
+ \sin\frac{\theta}{2}e^{i\varphi}|g_{2}\rangle,\nonumber\\
|E_{+}\rangle& = &\sin\frac{\theta}{2}e^{-i\varphi}|g_{1}\rangle
- \cos\frac{\theta}{2}|g_{2}\rangle + |e\rangle, \nonumber\\
|E_{-}\rangle&=&\sin\frac{\theta}{2}e^{-i\varphi}|g_{1}\rangle -
\cos\frac{\theta}{2}|g_{2}\rangle -|e\rangle,
\end{eqnarray}
and the corresponding eigenvalues are $E_{0}=0$, $E_{+}=G_{0}(t)$,
and $E_{-}=-G_{0}(t)$, respectively. In the dressed state
representation, we can get  the bright state
$|b\rangle=\sin\frac{\theta}{2}e^{-i\varphi}|g_{1}\rangle -
\cos\frac{\theta}{2}|g_{2}\rangle$ and  the dark state
$|d\rangle=\cos\frac{\theta}{2}|g_{1}\rangle +
\sin\frac{\theta}{2}e^{i\varphi}|g_{2}\rangle$. The bright state
couples to the excited state $|e\rangle$ and the dark state
decouples from the state $|e\rangle$.

\begin{figure}[!ht]
\begin{center}
\includegraphics[width=8.0cm,angle=0]{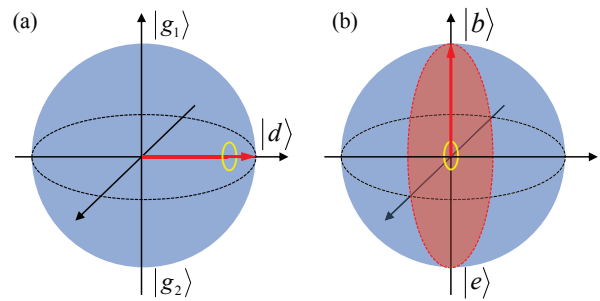}
\caption{ Two geometric dynamics process pictures for the
non-adiabatic quantum state transfer process. (a) Bloch sphere for
the evolution of the dark state vector; (b) Bloch sphere for the
evolution of the bright state vector. Under the cyclic evolution and
the parallel-transport condition, the dark state and the bright
state acquire the geometric phases of 0 and $\pi$, respectively.
}\label{fig2}
\end{center}
\end{figure}

\section{Universal single-qubit non-adiabatic
holonomic quantum gates in an optomechanical system}  \label{sec3}

To implement the single-qubit gates based on the non-adiabatic
geometric dynamics in an optomechanical system, two conditions
\cite{SjoqvistNJP2012} should be satisfied. First, one should make
$\alpha(\tau)=\pi$ with $\int^{t}_{0} G_{0}(t)dt=\alpha(t)$ to
ensure the states undergo a cyclic evolution. Second, one should
make the parallel-transport condition with
$\hat{H}_{ij}=\langle\psi_{i}(t)|\hat{H}_{2}|\psi_{j}(t)\rangle=0$
to keep the zero dynamical phases. In this way, the total evolution
phases are the purely geometric phases. The bright and the dark states
evolve as
\begin{eqnarray}      \label{condition}
|\psi_{1}(t)\rangle&=&\hat{U}_{1}(t)|d\rangle=|d\rangle,\nonumber\\
|\psi_{2}(t)\rangle&=&e^{i\alpha(t)}\hat{U}_{1}(t)|b\rangle\nonumber\\
&=&e^{i\alpha(t)}\left[\cos\alpha(t)|b\rangle\!-\!
i\sin\alpha(t)|e\rangle\right],
\end{eqnarray}
where the evolution operator is
$\hat{U}_{1}(t)=\exp\left(-i\int^{t}_{0}\hat{H}_{2}dt\right)$. The
inserted factor $e^{i\alpha(t)}$ is used to ensure the cyclic
evolution $|\psi_{2}(0)\rangle=|\psi_{2}(\tau)\rangle$ in the
projective Hilbert space. According to  Eq.(\ref{condition}), one
can derive that the accumulated  purely geometric phases during the
evolution process of the dark state and the bright state are $0$ and
$\pi$, respectively. As shown in Fig.~\ref{fig2}, the dark state
keeps unchange in the evolution process under the driving of the
Hamiltonian $H_{2}$ with the basis $\{|g_{1}\rangle,|g_{2}\rangle\}$
and the bright state evolves along the longitude with the basis
$\{|b\rangle,|e\rangle$\}.

Changing the dark-bright basis into the subspace spanned by
$\{|g_{1}\rangle,|g_{2}\rangle\}$, one makes a transformation of
coordinates with the form
\begin{eqnarray}        \label{eqgg}
|\xi_{1}(t)\rangle&=&\sin\frac{\theta}{2}e^{i\varphi}|\psi_{2}(t)\rangle
+ \cos\frac{\theta}{2}|d\rangle,\nonumber\\
|\xi_{2}(t)\rangle&=&-\cos\frac{\theta}{2}|\psi_{2}(t)\rangle
+ \sin\frac{\theta}{2}e^{-i\varphi}|d\rangle.
\end{eqnarray}
The above computational states satisfy
$|\xi_{l}(0)\rangle\!=|\xi_{l}(\tau)\rangle =|g_{l}\rangle$
$(l=1,2)$ to ensure the cyclic evolution after the above
transformations. The non-adiabatic holonomic dynamics can be
described by
$\hat{U}(\theta,\varphi)=\hat{T}\exp\left[i\int^{\tau}_{0}\hat{\mathcal{A}}dt\right]$,
where $\hat{T}$ is the time-ordering operator
\cite{SjoqvistNJP2012}. The matrix
$\hat{\mathcal{A}}_{ij}=i\langle\xi_{i}(t)|\partial_{t}|\xi_{j}(t)\rangle$
is given by
\begin{eqnarray}        \label{eqgg}
\hat{\mathcal{A}}=\dot{\alpha}(t)\left[
\begin{array}{cc}
-\sin^{2}\frac{\theta}{2}&\ \ \  e^{-i\varphi}\sin\frac{\theta}{2}\cos\frac{\theta}{2}\\
e^{i\varphi}\sin\frac{\theta}{2}\cos\frac{\theta}{2}&\ \ \ -\cos^{2}\frac{\theta}{2}\\
\end{array}
\right].
\end{eqnarray}
Therefore, one can  obtain the evolution operator with
\begin{eqnarray}        \label{eqgg}
\hat{U}(\theta,\varphi)=\left[
\begin{array}{cc}
\cos\theta&\sin\theta e^{-i\varphi}\\
\sin\theta e^{i\varphi}&-\cos\theta
\end{array}
\right],
\end{eqnarray}
where  $\theta$ and $\varphi$ are the corresponding  parameter
values in the Bloch sphere. By changing the different values of
coupling strength, i.e, $\theta$ and $\varphi$, one can get the NOT
gate, rotation gate, and Hadamard gate with
$(\theta,\varphi)=(\frac{\pi}{2},0)$,
$(\frac{\pi}{2},\frac{\pi}{8})$, and $(\frac{\pi}{4},0)$,
respectively \cite{DuanNature2014}. One can realize a phase gate by
the combination of $U(\frac{\pi}{2},\frac{\pi}{4})$ and
$U(\frac{\pi}{2},0)$
\begin{eqnarray}        \label{eqgg}
\left[
\begin{array}{cc}
\!\!0&\!\!e^{-i\frac{\pi}{4}}\\
\!\!e^{i\frac{\pi}{4}}&\!\!0\\
\end{array}
\right]\left[
\begin{array}{cc}
0&1\\
1&0\\
\end{array}
\right]=\left[
\begin{array}{cc}
\!\!e^{-i\frac{\pi}{4}}&\!\!\!\!0\\
\!\!0&\!\!\!\!e^{i\frac{\pi}{4}}\\
\end{array}
\right].
\end{eqnarray}
And the phase-flip gate can be given with  $\theta=0$ and
$\varphi=0$
 \begin{eqnarray}        \label{eqgg}
\hat{U}(0,0)=\left[
\begin{array}{cc}
1&0\\
0&-1
\end{array}
\right].
\end{eqnarray}

With these gates, one can obtain a set of universal single-qubit
gates which are based on the subspace spanned by
$\{|g_{1}\rangle,|g_{2}\rangle\}$. Besides, the NOT gate and
Hadamard gate can be applied in the optomechanics.

\begin{figure}[!ht]
\begin{center}
\includegraphics[width=7.5cm,angle=0]{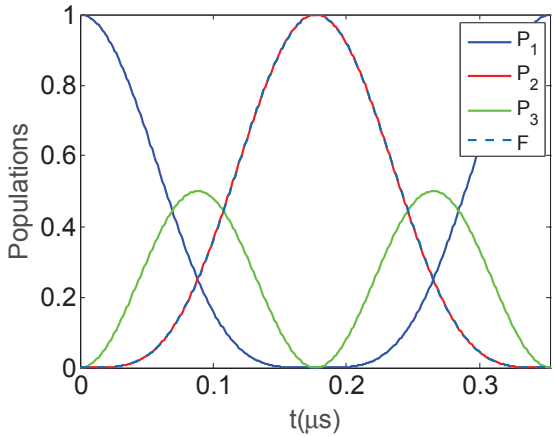}
\caption{ The simulation of the populations and the fidelity of the
NOT gate.  $P_{1}$, $P_{2}$, and $P_{3}$  represent the qubit-state
populations of cavities 1,  2, and the phonon, respectively.  $F$ is
the fidelity of the   NOT gate. When the initial state is
$|g_{1}\rangle$ , the ideal  final state $|g_{2}\rangle$ can be
achieved when the evolution time meets the condition
$t=\frac{\pi}{G_{0}(t)}\approx0.177$ $\mu s$,  which means the
population of  cavity 1 can be transferred completely to that of
cavity 2. }\label{transfer}
\end{center}
\end{figure}

Now, we use the NOT gate and the Hadamard gate to accomplish the
quantum state transfer and the generation of the entanglement in the
optomechanical system, respectively. For the quantum state transfer,
with $\varphi=0$ and $\theta=\frac{\pi}{2}$, one can obtain a  NOT
gate given by
\begin{eqnarray}        \label{eqgg}
U(\frac{\pi}{2},0)=\left[
\begin{array}{cc}
0&1\\
1&0\\
\end{array}
\right].
\end{eqnarray}
If the initial quantum state is chosen with $|g_{1}\rangle$, one can
accomplish the quantum state transfer between two cavities under the
driving of the NOT gate described by
\begin{eqnarray}        \label{eqgg}
|g_{2}\rangle=U(\frac{\pi}{2},0)|g_{1}\rangle.
\end{eqnarray}
In this paper, we choose the coupling strengths with
$\frac{G_{1}(t)}{2\pi}=2$ MHz, $\frac{G_{2}(t)}{2\pi}=-2$ MHz, and
$\frac{G_{0}(t)}{2\pi}=2\sqrt{2}$ MHz to perform the quantum state
transfer. We calculate the variation of population and fidelities in
Fig.~\ref{transfer}. The fidelity is defined with
$F=\langle\psi_{ideal}|tr_{m}[\rho(t)]|\psi_{ideal}\rangle$, where
$|\psi_{ideal}\rangle$ represents the ideal final state. For the
quantum state transfer, $|\psi_{ideal}\rangle=|01\rangle$.
$tr_{m}[\rho(t)]$ represents a reduced density matrix, where the
mechanical oscillator degree of freedom has been removed by tracing.
One can find that when the time is
$t=\frac{\pi}{G_{0}}\approx0.177\mu s$, the complete population
inversion indicates the system achieves the quantum state transfer
successfully and the system satisfies the cyclic evolution very
well.

\begin{figure}[!ht]
\begin{center}
\includegraphics[width=7.5cm,angle=0]{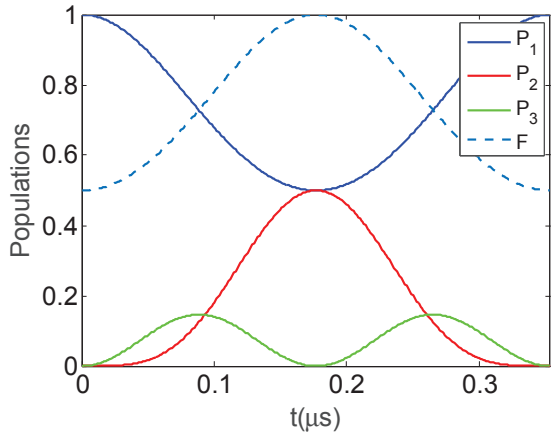}
\caption{ The simulation of the populations and the fidelity of the
Hadamard gate. $P_{1}$, $P_{2}$, and $P_{3}$   represent the
qubit-state populations of cavities 1, 2, and phonon,  respectively.
$F$ is the fidelity of the Hadamard gate. With the initial state as
$|g_{1}\rangle$, we realize the quantum entanglement between the two
cavities at the time $t=\frac{\pi}{G_{0}(t)}\approx0.177\mu s$.
}\label{entanglement}
\end{center}
\end{figure}

Also, one can generate a discrete variable entangled state
$|\psi_{ideal}\rangle=(|10\rangle + |01\rangle)/\sqrt{2}$ between
two cavities by choosing $\varphi=0$ and $\theta=\frac{\pi}{4}$ to
construct a Hadamard gate. The process is given by
\begin{eqnarray}        \label{eqgg}
|\psi_{ideal}\rangle|0\rangle_{b}=U(\frac{\pi}{4},0)|g_{1}\rangle.
\end{eqnarray}
Here, the Hadamard gate is given by \begin{eqnarray}        \label{eqgg}
U(\frac{\pi}{4},0)=\frac{1}{\sqrt{2}}\left[
\begin{array}{ccc}
1&1\\
1&-1\\
\end{array}
\right].
\end{eqnarray}
With the parameters $\frac{G_{1}(t)}{2\pi}\approx1.0824$ MHz,
$\frac{G_{2}(t)}{2\pi}\approx-2.6131$ MHz, and
$\frac{G_{0}(t)}{2\pi}\approx2.8284$ MHz, one performs the process
of entanglement generation in Fig.~\ref{entanglement}. When the
evolution time $t=\frac{\pi}{G_{0}(t)}\approx0.177\mu s$ is
satisfied, the fidelity arrives $100\%$, and the state becomes
$|\psi_{ideal}\rangle$.

\section{simulations and fidelities} \label{sec4}

To evaluate the performance of the SQNAHQGs, we calculate the fidelities of the NOT
gate and the Hadamard gate. With dissipation, the dynamics of process can be calculated
by the master equation with the Lindblad form given by
\begin{eqnarray}        \label{eq2}
\frac{d\hat{\rho}}{dt}\!=\!i[\hat{\rho},\hat{H}_{1}]\!+\!\kappa_{1}\hat{L}[\hat{a}_{1}]\hat{\rho}\!
+\!\kappa_{2}\hat{L}[\hat{a}_{2}]\hat{\rho}\!+\!\gamma_{m}\hat{D}[\hat{b}_{m}]\hat{\rho},
\end{eqnarray}
where  $\gamma_{m}$, $\kappa_{1}$ and $\kappa_{2}$ represent the
mechanical damping rate, the decay rates of the cavities 1 and 2,
respectively. $\hat{L}[\hat{o}]\hat{\rho}=(2\hat{o}\hat{\rho}
\hat{o}^{\dag}-\hat{o}^{\dag}\hat{o}\rho-\hat{\rho}
\hat{o}^{\dag}\hat{o})/2$.
$\hat{D}[\hat{o}]\hat{\rho}=(n_{th}+1)(2\hat{o}\hat{\rho}
\hat{o}^{\dag}-\hat{o}^{\dag}\hat{o}\hat{\rho}-\hat{\rho
}\hat{o}^{\dag}\hat{o})/2+n_{th}(2\hat{o}^{\dag}\hat{\rho}
\hat{o}-\hat{o}\hat{o}^{\dag}\hat{\rho}-\hat{\rho}
\hat{o}\hat{o}^{\dag})/2$, where $n_{th}$ is the thermal phonon
number of the environment. $\hat{\rho}$ is the density operator and
$\hat{H}_{1}$ is the Hamiltonian of the optomechanical system.

Here, we choose the  parameters $\kappa=\kappa_{1}=\kappa_{2}$ and
$\gamma_{m}$ with the range $\kappa=[0.31,6.28]\times10^{-1}$ MHz
and $\gamma_{m}=[1.88,35.81]\times10^{-3}$ MHz, respectively.
$n_{th}=100$. The  frequencies of the cavities 1,   2,  and the
mechanical oscillator are $\omega_{1}/2\pi\sim 100$ THz,
$\omega_{2}/2\pi\sim 100 $ THz, and $\omega_{m}/2\pi\sim 1$ MHz,
respectively. For the NOT gate, the influence induced by different $\kappa $ and $%
\gamma _{m}$\ on the fidelity of the quantum state transfer is shown in Fig.
5, and the maximum and minimum fidelities are 0.96 and 0.56, respectively.
The fidelity is inversely proportional to $\kappa $ and $\gamma _{m}$. For
the Hadamard gate, the maximum and the minimum fidelities become,
accordingly, 0.97 and 0.65. In the both cases, the fidelity tends to
decrease monotonously with respect to $\kappa $ and $\gamma _{m}$. The
higher the damping of the mechanical oscillator or the cavity mode is, the
lower the fidelity we can obtain will be. Therefore, the preparing of the
high quality optomechanical system is helpful for implementing universal
single-qubit holonomic gates with a high fidelity.

\begin{figure}[!ht]
\begin{center}
\includegraphics[width=7.5cm,angle=0]{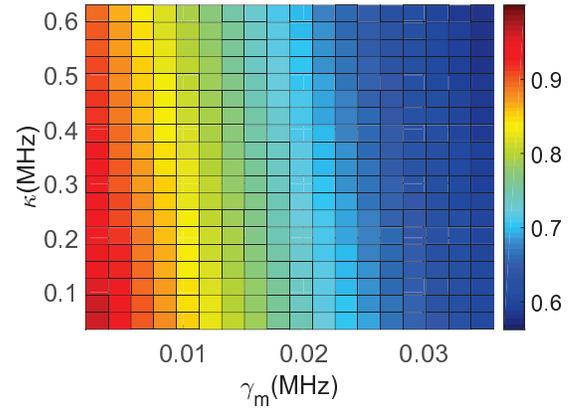}
\caption{ The fidelity of the NOT gate vs the decays of the cavities $\kappa
_{i}$ ($i=1,2$) and the dissipation of the mechanical oscillator $\gamma _{m}$. Here we choose
$\frac{G_{1}}{2\pi}=2$ MHz, $\frac{G_{2}}{2\pi}=-2$ MHz.
$\kappa=\kappa_{1}=\kappa_{2}$ and $\gamma_{m}$ with the range
$\kappa=[0.31,6.28]\times10^{-1}$ MHz and
$\gamma_{m}=[1.88,35.81]\times10^{-3}$ MHz,
respectively.}\label{transfer2}
\end{center}
\end{figure}

\begin{figure}[!ht]
\begin{center}
\includegraphics[width=7.5cm,angle=0]{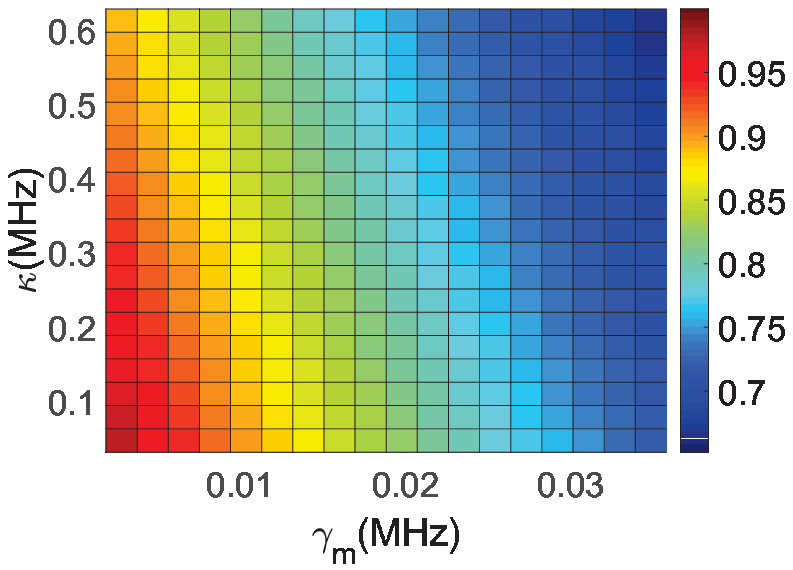}
\caption{ The fidelity of the Hadamard gate vs the decays of the cavities $\kappa
_{i}$ ($i=1,2$) and the dissipation of the mechanical resonator $\gamma _{m}$.
Here we choose $\frac{G_{1}}{2\pi}\sim1.0824$ MHz,
$\frac{G_{2}}{2\pi}\sim-2.6131$ MHz.
$\kappa=\kappa_{1}=\kappa_{2}$ and $\gamma_{m}$ with the range
$\kappa=[0.31,6.28]\times10^{-1}$ MHz and
$\gamma_{m}=[1.88,35.81]\times10^{-3}$ MHz,
respectively.}\label{entanglement2}
\end{center}
\end{figure}

\section{summary} \label{sec5}

In summary, we have proposed a prototype of the universal single-qubit quantum gates based on an optomechanical system working with the non-adiabatic geometric phases. We have shown its typical application by changing the different coupling strengths to get various noncommute quantum gates, such as Not gates, phase gates and Hadamard gates, and apply these gates for achieving quantum state transfer and the two-cavity mode entanglement generation in the optomechanical system. The result has shown that the quantum gates can have high fidelity against the negative influence of their dissipative environment. Our scheme is of all the good properties of the NHQC based on a quantum system and can be extended into other hybrid optomechanical quantum systems.\\

\section*{ACKNOWLEDGMENT}

This work is supported by the National Natural Science Foundation of
China under Grants No. 11654003, No. 11174040, No. 61675028, No.
11474026, and No. 11674033,  the Fundamental Research Funds for the
Central Universities under Grant No. 2015KJJCA01 and No. 2017TZ01, and the National
High Technology Research and Development Program of China under
Grant No. 2013AA122902.


\end{document}